\documentclass[12pt,preprint,superscriptaddress,showpacs]{revtex4}

\usepackage{mathrsfs}
\usepackage{amsmath}
\usepackage{amssymb}
\usepackage{graphicx}
\newcommand{\msl}{\mathscr{L}}

\begin{document}

\title{Quantum Electrodynamical Effects in Dusty Plasmas}

\author{M.\ Marklund}
\affiliation{Department of Physics, Ume{\aa} University, SE--901 87
  Ume{\aa}, Sweden}
\affiliation{Centre for Fundamental Physics,
  Rutherford Appleton Laboratory,
  Chilton Didcot, Oxfordshire, OX11 0QX, UK}
  
\author{L.\ Stenflo} 
\affiliation{Department of Physics, Ume{\aa} University, SE--901 87
  Ume{\aa}, Sweden}
  
\author{P.K.\ Shukla} 
\affiliation{Department of Physics, Ume{\aa} University, SE--901 87
  Ume{\aa}, Sweden}
\affiliation{Centre for Fundamental Physics,
  Rutherford Appleton Laboratory,
  Chilton Didcot, Oxfordshire, OX11 0QX, UK}

\author{G.\ Brodin}
\affiliation{Department of Physics, Ume{\aa} University, SE--901 87
  Ume{\aa}, Sweden}
\affiliation{Centre for Fundamental Physics,
  Rutherford Appleton Laboratory,
  Chilton Didcot, Oxfordshire, OX11 0QX, UK}
  
\date{\today}

\begin{abstract}
A new nonlinear electromagnetic wave mode in a magnetized dusty plasma is
predicted. Its existence depends on the interaction of an intense circularly 
polarized electromagnetic wave with a dusty plasma, where quantum
electrodynamical photon--photon scattering is taken into account. 
Specifically, we consider a dusty electron--positron--ion plasma, and show 
that the propagation of the new mode is admitted. It could
be of significance for the physics of supernova remnants and in neutron star formation. 
\end{abstract}


\maketitle

\section{Introduction}

Dusty plasmas play an important role in planetary as well as in astrophysical 
systems \cite{Shukla-Mamun,Horanyi-etal,Okamoto-etal}. However, many prominent 
astrophysical environments in which dusty plasmas can
be found are highly energetic and strongly magnetized, such as supernovae, neutron 
stars, and young stellar objects (YSO). There are important new wave modes is such 
dusty plasmas \cite{Stenflo-Shukla}, and the dust grain speeds may even become 
relativistic \cite{Bingham-Tsytovicha,Tsytovich-Bingham}. Supernovae, in which 
extreme energy scales are reached, allow for many exotic phenomena. In these violent events, 
neutron stars, YSO and pulsars are formed. Quantum electrodynamics (QED) is an indispensable 
explanatory model for much of the observed phenomena around these highly condensed objects. 
Scattering of photons off photons is predicted by QED, and it can be a prominent component 
of pulsar physics, since pulsars offer the necessary energy scales for such scattering to occur. 
Related to the scattering of photons is the concept of photon splitting in 
strong magnetic fields \cite{Adler}. It has been suggested that this effect 
could be important in explaining the radio silence of magnetars 
\cite{Kouveliotou-etal,Baring-Harding}. 
Moreover, the propagation of intense electromagnetic waves in plasmas gives 
rise to many interesting phenomena \cite{Mendonca}. 
The existence of
new plasma modes due to QED effects has been noted recently 
\cite{Stenflo-etal,Marklund-etala,Marklund-etalb}. In 
the present Letter, we point out the existence of a different electromagnetic 
wave that may exist  due to the interaction of charged dust and photons with 
the nonlinear quantum vacuum. 

\section{Basic equations}

First, let us consider the classical propagation of a circularly polarized electromagnetic 
wave in a multi-component plasma embedded in an external magnetic field $B_0 \hat {\bf z}$,
where $B_0$ is the strength of the external magnetic field and $\hat {\bf z}$ is the
unit vector along the $z$ axis. The dispersion relation for a circularly 
polarized wave with frequency $\omega$ and wavevector $k\hat{\mathbf{z}}$ 
is then given by (see e.g. Refs.\ \cite{Stenflo-Tsintsadze} and \cite{Shukla-etala})
\begin{equation}\label{eq:disprel}
  \frac{k^2c^2}{\omega^2} = 1 
  - \sum_j\frac{\omega_{pj}^2}{\omega(\omega\gamma_j 
    \pm \omega_{cj})}
\end{equation}
where the sum is over all particle species $j$,
$\omega_{cj} = q_jB_0/m_{0j}$ and 
$\omega_{pj} = (n_{0j} q_j^2/\epsilon_0 m_{0j})^{1/2}$ 
is the gyrofrequency and plasma frequency, respectively,
$\gamma_j$ is the relativistic gamma factor of species $j$, and
the upper (lower) sign in the denominator corresponds to left (right) hand
polarization. 
Here, $n_{0j}$ denotes particle
density in the laboratory frame, $q_j$ charge, $m_{0j}$ particle rest mass,
and $c$ the speed of light in vacuum. 
The plasma is assumed to consist of an admixture of electrons ($e$), 
positrons ($p$), ions ($i$), and charged dust particles ($d$). 
We note that the dispersion relation (\ref{eq:disprel}) has been derived without
applying the usual small amplitude linearization.
For a given $k$, the polarization is determined by the sign of $\omega$.   
In the background state, the plasma is close to equilibrium, and we 
have approximate charge neutrality, $\sum_jq_j n_{0j} \approx 0$. Thus,
for strongly magnetized relativistic electrons, positrons and ions, a 
low-frequency ($\omega \ll kc$ and $\omega\gamma_{e,p,i} \ll \omega_{ce,cp,ci}$) 
wave will satisfy the dispersion relation 
\begin{equation}
\label{eq:classical}
  \frac{k^2c^2}{\omega^2} \approx 
  - \frac{\omega_{pd}^2\gamma_d}{\omega_{cd}(\mp\omega\gamma_d - \omega_{cd})} ,
\end{equation}
which is appropriate for large amplitude electromagnetic waves in supernova. 
Here, $\gamma_d = (1 + \nu^2)^{1/2}$, where
\begin{equation}
  \nu^2 = \left( \frac{Z_deE_0}{cm_{od}} \right)^2%
    \frac{1 + \nu^2}{\left[ \omega(1 + \nu^2)^{1/2} \pm \omega_{cd} \right]^2} ,
\end{equation}
gives the relativistic gamma factor for dust with charge $q_d = Z_de$, where $e$
is the magnitude of the electron charge.
Henceforth we will allow $\omega < 0$, thus incorporating the 
left hand polarization as negative frequencies, and removing the $\mp$ in Eq.\ 
(\ref{eq:classical}).

Next, we investigate the effects of the nonlinear quantum vacuum
on the dispersion relation (\ref{eq:classical}) of a circularly polarized electromagnetic wave 
in a dusty magnetoplasma.

The weak field theory of photon--photon scattering can be formulated in terms of the 
effective Lagrangian density
\begin{equation}
  \msl = \msl_0 + \msl_{\mathrm{HE}},
\end{equation}
where $\msl_0 = -\tfrac{1}{4}\epsilon_0F_{ab}F^{ab} =
\tfrac{1}{2}\epsilon_0({\bf E}^2 - c^2{\bf B}^2)$ is the classical free
field Lagrangian, and 
\begin{equation}\label{eq:lagrangian}
  \msl_{\mathrm{HE}} = \kappa\epsilon_0^2\left[4\left(
  \tfrac{1}{4}F_{ab}F^{ab}\right)^2 +
  7\left( \tfrac{1}{4}F_{ab}\widehat{F}^{ab} \right)^2 \right],
\end{equation}
is the Heisenberg--Euler correction \cite{Heisenberg-Euler,Schwinger},
where $\widehat{F}_{ab} = \tfrac{1}{2}\epsilon_{abcd}F^{cd}$, and
$\tfrac{1}{4}\widehat{F}_{ab}F^{ab} = - c{\bf E}\cdot{\bf B}$.
Here, $\kappa \equiv 2\alpha^2\hbar^3/45m_{0e}^4c^5 \approx 1.63\times
10^{-30}\, \mathrm{m}\mathrm{s}^{2}/\mathrm{kg}$, $\alpha$ is
the fine-structure
constant, and $\hbar$ is the Planck constant.
With $F_{ab} = \partial_aA_b - \partial_bA_a$, $A^b$ being the four-potential, 
we obtain (see, e.g.\ Ref.\ \cite{Shukla-etalb})
\begin{equation}
  \partial_bF^{ab} = 2\epsilon_0\kappa\partial_b\left[ 
  (F_{cd}F^{cd})F^{ab} 
  + \tfrac{7}{4}(F_{cd}\widehat{F}^{cd})\widehat{F}^{ab} \right]
  + \mu_0 j^a ,
  \label{eq:maxwell}
\end{equation}
where $j^a$ is the four current.

For a magnetic field-aligned circularly polarized electromagnetic wave 
with the electric field $\mathbf{E}_0 = E_{0} 
(\hat{\mathbf{x}} + i\hat{\mathbf{y}})\exp(ikz
  - i\omega t)$ , the invariants satisfy
 $ F_{cd}F^{cd} = -2E_0^2( 1 - k^2c^2/\omega^2) 
    + 2c^2B_0^2$ and 
$  F_{cd}\widehat{F}^{cd} = 0. $
Thus, Eq.\ (\ref{eq:maxwell}) can be written as
\begin{equation}
  \Box A^a = -4\epsilon_0\kappa\left[ E_0^2\left( 1 - \frac{k^2c^2}{\omega^2}\right) 
    - c^2B_0^2 \right]\Box A^a + \mu_0 j^a ,
  \label{eq:wave}
\end{equation}
in the Lorentz gauge, and $\Box = \partial_a\partial^a$.
For right-hand circularly polarized electromagnetic waves 
propagating in a magnetized cold dusty pair-ion plasma, the matter effects can 
be absorbed in the wave operator on the left-hand side by the replacement
$ \Box \rightarrow -D(\omega, k)$, where $D$ is the plasma dispersion function, 
given by (see, e.g.\ Refs.\ \cite{Stenflo,Stenflo-Tsintsadze,Stenflo-Shukla}, 
and Eq.\ (\ref{eq:classical}))
\begin{equation}
  D(\omega,k) = k^2c^2 
  + \frac{\omega^2\omega_{pd}^2\gamma_d}{\omega_{cd}(\omega\gamma_d 
    - \omega_{cd})} .
\label{eq:dispersionfunction}   
\end{equation}
We stress that the application of Eq. (\ref{eq:lagrangian}) 
requires field strengths smaller than the Schwinger 
critical field $E_S = m_{0e}^2c^3/e\hbar \sim 10^{18}$ V/m [see, e.g.\ Ref.\ 
\cite{Schwinger}). 
Looking for low frequency modes, the dispersion relation obtained from 
Eqs.\ (\ref{eq:wave}) and (\ref{eq:dispersionfunction}) then reads
\begin{eqnarray}
    \frac{k^{2} c^{2} }{\omega^{2} } \approx \frac{4\alpha}{45\pi}\left[\left( 
    \frac{E_0}{E_S}  \right)^2 \frac{k^{2} c^{2} }{\omega^{2} } 
    + \left(\frac{cB_0}{E_S}\right)^2 \right]\frac{k^2c^2}{\omega^2}  
    - \frac{\omega_{pd}^2\gamma_d}{\omega_{cd}(\omega\gamma_d - \omega_{cd})} .  
\label{eq:transverse2}
\end{eqnarray}
Equation (\ref{eq:transverse2}) is the main result of
our calculations. Due to their large mass, the dust particles can under most circumstances be 
considered non-relativistic, i.e.\ $\gamma_d \approx 1$ 
\cite{Bingham-Tsytovicha,Tsytovich-Bingham}. We note the following for the dispersion relation 
(\ref{eq:transverse2}): In the classical case, according to Eq.\ (\ref{eq:classical})
we must have $\omega\gamma_d - \omega_{cd} < 0$. This requirement is
obviously not necessary for the QED case, where $\omega\gamma_d - \omega_{cd} \geq 0$
is allowed. 
Thus, new low frequency waves are expected in the strong
field regime. 

\section{Analysis of the dispersion relation}

In order to gain a better understanding of the dispersion relation 
(\ref{eq:transverse2}) we introduce the dimensionless variables and parameters 
$\Omega = \omega/\omega_{cd}$, $K = kc/\omega_{pd}$,
$\Omega_p = \omega_{pd}/\omega_{cd}$, 
$\sigma = (4\alpha/45\pi)(cB_0/E_S)^2$, and $\delta = E_0/cB_0$. The dispersion
relation (\ref{eq:transverse2}) then reads
\begin{equation}\label{eq:dimless}
  \left(\frac{K}{\Omega}\right)^2 = \sigma\left[ 
    1 + \delta^2\Omega_p^2\left(\frac{K}{\Omega}\right)^2 
      \right]\left(\frac{K}{\Omega}\right)^2 + \frac{\gamma_d}{1 
         - \Omega\gamma_d} ,
\end{equation}
where the relativistic gamma factor can be found from the relation
\begin{equation}\label{eq:gamma}
  \frac{\gamma_d^2 - 1}{\gamma_d^2} = \frac{\delta^2}{(1 - \Omega\gamma_d)^2} .
\end{equation}

We may now discern between the following cases:

(a) $\gamma_d \sim 1$: If we let $\sigma \rightarrow 0$, i.e. neglect the QED
correction, we only have mode propagation for $\Omega -1  
< 0$. When $\sigma \neq 0$, this is no longer a necessary condition for 
mode propagation, and we can therefore have $\Omega - 1 > 0$. A special case here
is the strongly magnetized dusty plasma, $cB_0 \gg E_0$.  
Equation (\ref{eq:transverse2}) then yields
\begin{equation}
  (1 - \sigma)\left(\frac{K}{\Omega}\right)^2 = \frac{1}{1 - \Omega} ,
\end{equation}
for which the QED effects act only as a small correction.

(b) $\gamma_d \gg 1$: From Eqs.\ (\ref{eq:dimless}) and (\ref{eq:gamma})
we obtain the dispersion relation
\begin{equation}
  \left(\frac{K}{\Omega}\right)^2 = \sigma\left[ 
    1 + \delta^2\Omega_p^2\left(\frac{K}{\Omega}\right)^2 
      \right]\left(\frac{K}{\Omega}\right)^2 + \frac{-\delta \pm 1}{\Omega\delta} .
\end{equation}
In the limit $\sigma \rightarrow 0$, we have a single mode given by
\begin{equation}
  \left(\frac{K}{\Omega}\right)^2 = \frac{-\delta + 1}{\Omega\delta} .
\end{equation}
When $\sigma \neq 0$, we
have further completely distinct modes as given by the dispersion relation
\begin{equation}\label{eq:new1}
  \left(\frac{K}{\Omega}\right)^2 = \sigma\left[ 
    1 + \delta^2\Omega_p^2\left(\frac{K}{\Omega}\right)^2 
      \right]\left(\frac{K}{\Omega}\right)^2 - \frac{1 + \delta}{\Omega\delta} .
\end{equation}
As we will demonstrate, Eq.\ (\ref{eq:new1}) has new branches which is not present in
the classical theory. Moreover, it will be further pronounced for 
low frequency waves, and we therefore expect the QED effects to
play an important role in the propagation of such modes. In the limit of large $\delta$, 
i.e.\  weak external magnetic field $B_0$, we can simplify the dispersion relation to
\begin{equation}\label{eq:new2}
  \left(\frac{K}{\Omega}\right)^2 = \sigma\delta^2\Omega_p^2\left(\frac{K}{\Omega}\right)^4 
     - \frac{1}{\Omega} ,
\end{equation}
in order to further elucidate the character of the new QED mode. 
In the limit $\sigma \rightarrow 0$, we obtain the whistler-like
dispersion relation $\Omega = -K^2$ (see Fig.\ 1).

\section{Applications and conclusions}

In astrophysical environments, such as supernovae, it is believed that charged dust grains 
can affect the plasma dynamics in crucial ways 
\cite{Bingham-Tsytovichb,Popel-Tsytovich,Sedlmayr}. 
It has also been argued that supernovae can be a significant source of dust particles 
\cite{Dunne-etal}, although this claim is under debate \cite{Krause-etal}.
At the same time the plasma may be highly magnetized, thus altering the behavior of the 
dusty plasma significantly \cite{Sato}.  In fact, the fields can reach amplitudes
such that the vacuum will be excited, 
making it nonlinear due to QED effects.  
In this perspective, it is thus natural that the mode structure presented here is more complex 
than that of linear theory.
In Fig.\ 1, we have plotted the dispersion curves, as given by Eq.\ 
(\ref{eq:new2}), for different values of the QED parameter $A$. 
We note that for $K$ larger than a critical value (which depends on $A$),
there are three branches of the dispersion relation (\ref{eq:new2}), 
in contrast to the single branch present in the classical limit. 
It should be stressed that the values
of $A$ used in Fig.\ 1 by no means require unrealistic
astrophysical conditions, as the following example will show: In interstellar 
clouds, the dust density is expected to be $n_{0d} \sim 10^{-1}\,\mathrm{m^{-3}}$ 
\cite{Mendis-Rosenberg}, with a typical dust mass $m_{0d} \sim 
10^6\,m_{\text{proton}} \sim 10^{-21}\,\mathrm{kg}$ \cite{Shukla-Mamun}. 
For the circularly polarized wave, we take the amplitude to satisfy
$E_0/E_S \approx 2\times10^{-5}$, a value appropriate at a distance of 
1\,AU from a supernova. In this case, a value 
$B_0 \sim 10^{-8}\,\mathrm{G}$ (well within the limits for such environments)
results in $A \sim 1$. For higher dust mass densities we note that even lower
wave intensities will give values of $A$ close to one. Such dense
dusty plasmas are likely to be found closer to the core of a supernova 
remnant. It should also be noted that we have neglected radiative-resonant
interactions on our new electromagnetic wave mode. However, in many cases such
interactions can dominate \cite{Tsytovich,Tsytovich-Bingham99}, 
thus altering the behavior of the
mode. This is a phenomenon that has to be considered in future research.
In addition, an important problem of the physics of supernova explosions is the
relation between the layer of dust behind the supernova shock and
the process of dust-wave condensation behind the shock wave front
\cite{Popel-etal}. Our new nonlinear wave mode may thus play a role from the
viewpoint of supernova shock propagation. 

We have presented an analysis regarding the propagation 
of intense circularly polarized waves in a magnetized dusty plasma.
It is found that corrections due to QED modify the mode structure,
giving rise to novel low frequency waves with no classical counterpart.     
Our main result is the dispersion relation (\ref{eq:transverse2}), in which 
the effects of charged dust and QED are taken into account.
The result should be of interest in astrophysical environments, such as supernova
remnants and interstellar clouds. 

\section*{Acknowledgements}
This research was partially supported by the Swedish Research Council 
through the contracts No. 621-2001-2274 and No. 621-2004-3217, as well 
as by the Deutsche Forschungsgemeinschaft through the Sonderforschungsbereich 591 and
the European Commission through the contract No. HPRN-CT-2001-00314.


\newpage

\begin{figure}
  \includegraphics[width=.9\columnwidth]{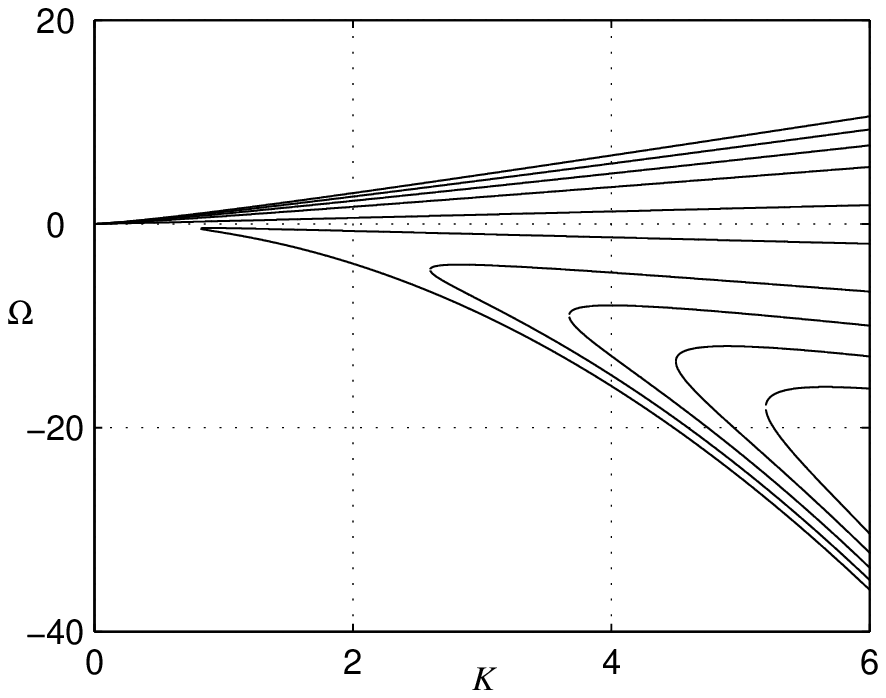}
  \caption{The normalized frequency $\Omega$ plotted as a function of the 
    normalized wavenumber $K$, as given by the relation (\ref{eq:new2}). 
    Variation of the parameter $A = \sigma\delta^2\Omega^2_p$ gives 
    the different solution curves. The curves in the upper and lower half 
    planes represents the roots of Eq.\ (\ref{eq:new2}). From the line $\Omega = 0$ outwards,
    the curves for the values $A = 0.1, 1, 2, 3$, and $4$ have been plotted. 
    We note that the modes for which $\Omega > 0$ (right-hand polarization) have no classical 
    counterpart. When $A \rightarrow 0$ we obtain $\Omega = -K^2$, which 
    represents a classical limit curve in the lower half plane.}
\end{figure}

\end{document}